# Development of a Mach-Zehnder Modulator Photonic Local Oscillator Source

Derek Y. Kubo, *Member, IEEE*, Ranjani Srinivasan, Hitoshi Kiuchi, Ming-Tang Chen

*Abstract*— **This paper describes the development of a photonic local oscillator (LO) source based on a 3-stage Mach-Zehnder modulator (MZM) device. The MZM laser synthesizer demonstrates the feasibility of providing the photonic reference LO for the Atacama Large Millimeter Array telescope located in Chile. This MZM approach to generating an LO by radio RF modulation of a monochromatic optical source provides the merits of wide frequency coverage of 4-130 GHz, tuning speed of about 0.2 seconds, and residual integrated phase noise performance of 0.3 degrees RMS at 100 GHz.**

## I. INTRODUCTION

THE generation of local oscillator (LO) sources for use in radio interferometry have traditionally made use of radio frequency (RF) components, such as YIG (Yttrium Iron Garnet) and Gunn oscillators, which can be phase locked to produce accurate frequency and low phase noise. This technology, however, does not easily scale to modern interferometers that are approaching near terahertz observing frequencies. One such example is the Atacama large millimeter array (ALMA) located in Chile. The ALMA telescope is designed to operate over a frequency range of 31-950 GHz in 10 bands, and to eventually consist of 66 12-m and 7-m antennas with heterogeneous spacing from 15-m to 15-km. The upper frequency of 950 GHz drives the design approach that utilizes optical distribution of a 27-122 GHz reference LO followed by a maximum of x9 multiplication within the antennas. The phase coherency of the final LO in each of the antennas dictates the phase coherency of the detected astronomical signal. The high operating frequency combined with the large disparity of optical fiber lengths to

Manuscript received March 04, 2013. This work was supported, in part, by the National Science Council, Taiwan, Republic of China, under the Atacama Large Millimeter Array – Taiwan project.

D. Y. Kubo is with the Academia Sinica Institute of Astronomy and Astrophysics, Hilo, HI 96720 USA (phone: 808-961-2926; fax: 808-961-2989; e-mail: dkubo@asiaa.sinica.edu.tw).

R. Srinivasan is with the Academia Sinica Institute of Astronomy and Astrophysics, Hilo, HI 96720 USA (e-mail: rsriniva@asiaa.sinica.edu.tw).

H. Kiuchi is the National Astronomical Observatory of Japan, Mitaka, Tokyo 181-8588 Japan (e-mail: hitoshi.kiuchi@nao.ac.jp).

M. T. Chen is with the Academia Sinica Institute of Astronomy and Astrophysics, Hilo, HI 96720 USA (e-mail: mchen@asiaa.sinica.edu.tw).

the antennas necessitates a reference LO with very low phase noise. The key instrument that dictates the overall central LO performance is the laser synthesizer used to generate a pair of optical tones separated by a tunable range of 27-34 GHz and 65-122 GHz. A set of 10 photomixers, one for each band, are used to generate the RF reference LOs within each of the antennas.

The ALMA project utilizes a laser synthesizer based on the beating of a phase locked slave laser with a master laser [1]. In this scheme, four sets of slave lasers, photomixers, and harmonic mixers are utilized to cover the photonic LO tuning range of 27-34 GHz and 65-122 GHz. To maintain LO coherency, the two lasers must be stable, spectrally pure, have narrow line widths, and be optically phase locked to each other. The practical design challenges in phase locking the slave lasers to the master laser and produce an LO with low phase noise was the impetus for the development of an alternate laser synthesizer based on a Mach Zehnder modulator (MZM) device.

This paper describes the development and performance of the MZM laser synthesizer for the ALMA project. The approach of generating a pair of optical sideband tones by modulating a monochromatic light source using a dual-parallel interferometric modulator is simple and eliminates the need to phase lock lasers. The two optical tones are generated from a common laser source and are inherently coherent, producing a low phase noise LO when the heterodyned beat note is detected by a photomixer.

## II. DESCRIPTION OF THE MZM DEVICE

### A. Brief History

The concept of using a MZM to generate a photonic LO dates back over three decades. Among the first attempts, [2], [3] demonstrated the use of an integrated lithium niobate waveguide as a traveling wave modulator. The modulation frequencies for those early experiments were limited to about 7.5 GHz. The next significant milestone for extending the bandwidth was achieved by [4] using an 18 GHz signal to modulate a standard distributed feedback (DFB) laser to produce a LO at 36 GHz. The precursor to the design of the MZM laser synthesizer discussed in this paper was successfully demonstrated by [5], [6]. A RF beat-note was produced at four times the modulating frequency to achieve LO signals ranging from 32-49 GHz. High spectral purity was demonstrated at the end of a 25-km spool of single-mode optical fiber that would later prove to be important for interferometry applications. Other frequency quadrupling schemes have been developed to increase the LO frequency range using a pair of cascaded MZM devices [7].

The first successful attempt at using a photonic LO in radio interferometric imaging was reported by [8], [9]. Test observations were conducted at the Smithsonian submillimeter array (SMA) telescope with the photonic LO incorporated into one element of a 5-antenna sub-array. Several quasars were detected and an ultra-compact HII region[1] was imaged.

The proposal of using a 3-stage MZM device to achieve high laser carrier suppression was explored by [10] as an alternative scheme to generate a photonic LO for the ALMA project. Further development of this concept is provided in [11]-[15].

### B. Description of the 3-stage MZM device

The key component of the laser synthesizer discussed in this paper is the MZM device. This device consists of a 3-stage MZM with the configuration and physical structure depicted in Fig. 1. The underlying principle for each cell is the varying index of refraction for the lithium niobate substrate as a function of applied electrode voltage [16]. Applying 0 V to a cell produces equal phase through the two arms and results in constructive interference at the output of the summing Y-node. Similarly, applying the half wave voltage, $V_\pi$, produces a relative phase difference of 180 degrees between the arms and results in destructive interference at the same Y-node. Note the applied electric field, $E_z$, in Fig. 1(b) is aligned in the z-direction and results in the largest change in the refractive index for the x-cut lithium niobate crystal [17], [18]. The $V_\pi$ voltage is specific to the device and can be made lower by increasing the length, L, of the cell dimension. The electric fields generated across the dielectric material are in opposing directions for the two optical waveguide paths. This push-pull configuration [11] reduces the required electrode voltage by a factor of two. It is important to note that the optical input signal must be polarized with the E-field aligned in the z-direction. Misalignment of the optical input polarization will have the effect of reduced modulation efficiency.

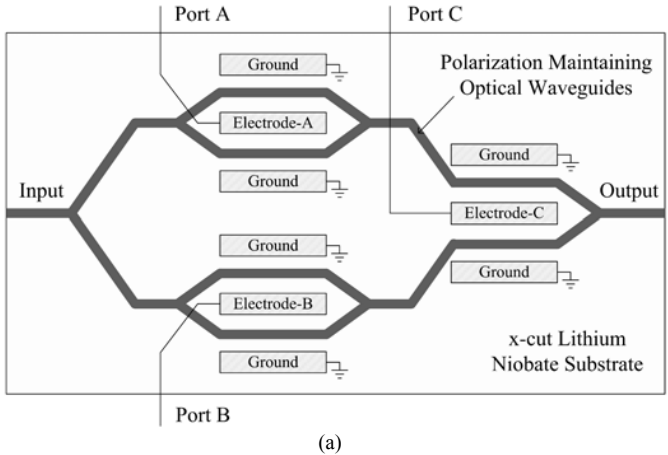

(a)

Port A Port C

Ground
Electrode-A
Ground
Polarization Maintaining
Optical Waveguides
Ground
Electrode-C
Input
Ground
Output
Ground
Electrode-B
Ground
x-cut Lithium
Niobate Substrate

Port B

HII region is an emission nebulae characterized by large amounts of ionized hydrogen.

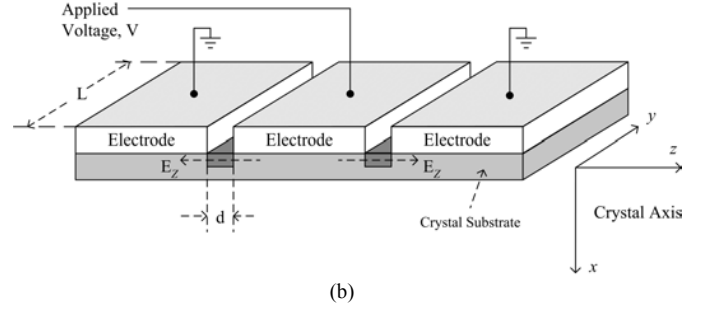

(b)

Fig. 1. 3-stage MZM device used in the laser synthesizer [11], [12]. (a) Ports labeled Input and Output are the optical interfaces. Ports A and B are electrical DC connections to fine adjust the optical delay through the upper and lower optical paths. Port C is an electrical connection provided with DC + RF to generate the desired optical tones space by either $2f_m$ (null-bias mode) or $4f_m$ (full-bias mode) centered about the optical carrier. (b) Cross-section view of x-cut type lithium niobate MZM cell. The two optical waveguide Y-branches are shown relative to the crystal axes. Note the opposing electric fields, $E_z$, producing a push-pull configuration.

### C. Optical Intensity Modulation Using an RF Signal

Application of an RF signal to port C of the MZM device will result in a modulated electric field that can be represented as

$$E(t) = 2E_m \sin(2\pi f_m t), \quad (1)$$

where $E_m$ and $f_m$ are the amplitude and frequency of the modulating signal, respectively, and a factor of 2 to accommodate the push-pull configuration. The optical output electric field intensity from the MZM [11] can be written as

$$E_{out} = \frac{E_0}{2} e^{2\pi i f_0 t} \left[ e^{+\left(\delta\sin(2\pi f_m t)+\frac{\theta_B}{2}\right)} + e^{-\left(\delta\sin(2\pi f_m t)+\frac{\theta_B}{2}\right)} \right]. \quad (2)$$

In the above equation the optical electric field with amplitude, $E_0/2$, and frequency, $f_0$, is phase modulated by a time varying sinusoidal field with frequency, $f_m$, and amplitude $\delta$. The term $\delta$ is defined as the phase modulation depth, $\delta = (\pi/V_\pi)V_m$, where $V_m = E_m d$. The phase modulation function across each individual arm can be represented by $+[\delta\sin(2\pi f_m t)+\theta_B/2]$ and $-[\delta\sin(2\pi f_m t)+\theta_B/2]$, respectively, where $\theta_B$ is the phase delay in radians imposed by the DC bias. Using the Bessel function identity (2) can be written as

$$E_{out} = \frac{E_0}{2} e^{2\pi i f_0 t} \sum_{n=-\infty}^{\infty} J_n(\delta) e^{2\pi i (nf_m t)} \left[ e^{\frac{i\theta_B}{2}} + (-1)^n e^{\frac{-i\theta_B}{2}} \right] \quad (3)$$

and

$$E_{out} = E_0 e^{2\pi i f_o t}\left[\cos\left(\frac{\theta_B}{2}\right)\sum_{n=-\infty}^{\infty}J_{2n}\left(\delta\right)e^{2\pi i\left(2nf_m t\right)}\right.$$
$$\left.+i\sin\left(\frac{\theta_B}{2}\right)\sum_{n=-\infty}^{\infty}J_{2n+1}\left(\delta\right)e^{2\pi i\left(\left(2n+1\right)f_m t\right)}\right]. \tag{4}$$

The optical output intensity can be obtained by constructing the product $|E_{out}|^2$. Retaining only the lower order terms in the Bessel expansion leads to

$$|E_{out}|^2 \simeq |E_0|^2\left[J_0^2\left(\delta\right)\cos^2\left(\frac{\theta_B}{2}\right)+2J_1^2\left(\delta\right)\sin^2\left(\frac{\theta_B}{2}\right)\right.$$
$$-4J_0\left(\delta\right)J_1\left(\delta\right)\sin\left(\frac{\theta_B}{2}\right)\cos\left(\frac{\theta_B}{2}\right)\sin\left(2\pi f_m t\right)$$
$$+2\cos\left(2\times 2\pi f_m t\right)\left\{2J_0\left(\delta\right)J_2\left(\delta\right)\cos^2\left(\frac{\theta_B}{2}\right)\right.$$
$$\left.\left.-J_1^2\left(\delta\right)\sin^2\left(\frac{\theta_B}{2}\right)\right\}\right] . \tag{5}$$

where $J_n(\delta)$ are the $n^{th}$ order Bessel function with argument $\delta$. The relative amplitudes for $J_n(\delta)$ for $n = 0, 1, 2,$ and 3 are plotted in Fig. 2 as a function of modulation depth, $\delta$. Note that for $\delta = 0$ (no modulation) only the pure laser carrier $J_0$ exists, and increasing $\delta$ generates even and odd harmonics $J_1$-$J_3$. Even or odd harmonics can be selected in (5) by changing the operating bias, $\theta_B$. Setting $\theta_B = 0$ radians provides only the original optical signal and the even harmonics. This condition is defined as operating in the full-bias mode. Conversely, setting $\theta_B = \pi$ radians gives us only the odd harmonics and suppresses the zero$^{th}$ and even harmonics. This condition is defined as operating in the null-bias mode. The two operational modes will be discussed in further detail in the next section.

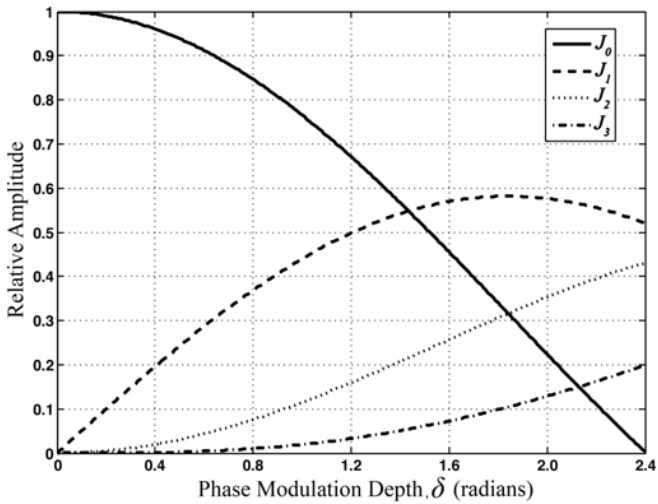

Fig. 2. Bessel functions of the first kind [11]. Relative amplitude of the zero$^{th}$ order Bessel function, $J_0$, with respect to the higher order Bessel functions $J_1$, $J_2$, and $J_3$ shown in the plot. Note that the actual power in the $n^{th}$ optical sidebands will be proportional to the square of the $n^{th}$ order Bessel function. $J_1$ and $J_3$ contribute to the amplitudes of the odd harmonics (retained in the null-bias mode), while $J_2$ contributes to the amplitude of the even harmonics (retained in the full-bias mode).

### D. Specific Design of the 3-Stage MZM Device

The MZM device used within the laser synthesizer utilizes a 3-stage x-cut lithium niobate modulator described earlier in Fig. 1. An optical polarizer is provided within the device to prevent processing of unpolarized or imperfectly polarized light. The sub-MZ stages are introduced as intensity trimmers to compensate for loss imbalances in the arms and within the main MZ stage. This allows for a much higher extinction ratio to be obtained by adjusting the bias to electrodes A and B. The half wave voltage value, $V_\pi$, at DC is approximately 5 V for all 3 ports. As will be described later, the actual value of $V_\pi$ varies as a function of modulation frequency. The DC bias to electrode C is applied along with the modulating RF signal via a bias-T device. This MZM device has a maximum modulation bandwidth of approximately 35 GHz and places an upper limit frequency of 70 GHz in null-bias mode, and 140 GHz in full-bias mode.

### III. MZM LASER SYNTHESIZER UNIT DESIGN

A block diagram of the MZM laser synthesizer is shown in Fig. 3. The unit is designed to accept a polarization maintained narrow line width laser at a wavelength of 1556.21-nm and a level of +13 dBm. The J1 input connector type is FC/APC and is physically keyed to align with the slow axis of the polarization maintaining (PM) fiber. All fibers and components within the unit are of PM type and are keyed to this convention. The optical input signal is sent through an isolator/polarizer device, IS1, which serves as isolation and to block any undesired polarization from entering MZ1. MZ1 is the 3-stage MZM device described in detail in the previous section. A combination of a low noise DC bias source and DC amplifier are used to bias ports A, B and C (Fig. 1(a)) of MZ1 over a range of ±10 V. Port C is DC biased through a bias-T device, BT1. A 13-31 GHz, +7 dBm RF input reference is provided to the unit via a 2.9-mm connector designated as J4. RF amplification is provided by power amplifier, AR1 (2-50 GHz), and delivers approximately +28 dBm of signal power to port C of MZ1 through BT1.

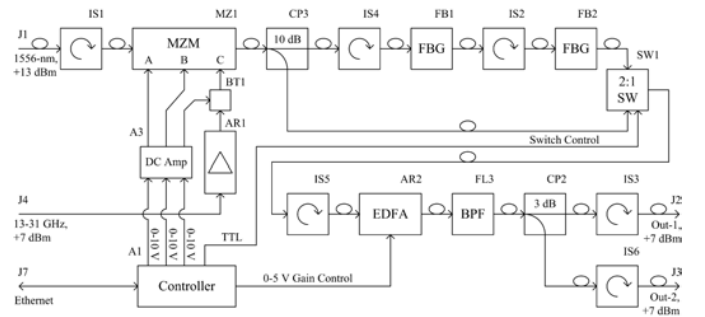

Fig. 3. Block diagram of MZM laser synthesizer. Device MZ1 provides RF modulation of the optical input signal. Amplifier AR1 provides +28 dBm input drive level to port C of MZ1. Optical switch SW1 selects between the upper full-bias and lower null-bias mode optical paths. An EDFA is used to compensate for optical component losses.

### A. Null-bias versus Full-bias Mode

The MZ1 device is operated in one of two modes, null-bias for 27-34 GHz, and full-bias for 65-122 GHz, and is controlled by the DC value applied to port C. The nominal values for null and full-bias modes are approximately 0 and +5 V, respectively. The optimal DC bias values, however, vary as a function of RF drive frequency and input power and therefore are derived empirically in 1 GHz increments of the reference input frequency and stored in two separate tables, one each for null and full-bias modes. Interpolation is used to set the MZ1 port C bias for any tuning frequency within the valid range. The fine trim DC bias values for ports A and B of MZ1 are empirically derived and kept static for both modes and all frequencies.

When MZ1 is set to null-bias mode, a two tone optical output is obtained and consists of the first harmonic lower sideband (LSB) and upper sideband (USB), namely $f_0 \pm f_m$. Therefore, the desired RF beat-note produced by the photomixer is twice the RF reference frequency or $2f_m$. The central laser at a vacuum wavelength of 1556.21-nm is suppressed by 25-30 dB. Fig. 4(a) represents the MZ1 optical output for null bias mode with 15.0 GHz applied to port-C. The two largest tones represent the desired 1st harmonic at a balanced level of +0.8 dBm and separated by 30 GHz. Note the 27 dB suppression of the 1556.21-nm laser. The undesired 3rd and 5th harmonic spurs are separated by 90 GHz (0.7270-nm) and 150 GHz (1.212-nm), respectively, with a worst case signal-to-spur ratio of 20 dB. Referring back to Fig. 3, in the null-bias mode the signal is coupled off by CP3 and selected by optical switch SW1. The 10 dB coupling ratio of CP3 serves to compensate for the higher tone output level in null-bias mode.

When MZ1 is set to full-bias mode, a three tone optical signal is obtained, which consists of the second harmonic LSB and USB, namely $f_0 \pm 2f_m$, and the central laser which is no longer suppressed. Neglecting the undesired laser tone, the desired RF beat-note produced by the photomixer is four times the RF reference frequency or $4f_m$. The spectral output of MZ1 in full-bias mode with 24 GHz applied to port-C is shown in Fig. 4(b). Note the large unsuppressed laser at +3 dBm. The desired 2nd harmonic tones are at a level of -11.0 dBm and are separated by 96 GHz. The undesired 1st and 3rd harmonic spurs are separated by 48 GHz (0.3878-nm) and 144 GHz (1.1633-nm), respectively, with a worst case signal-to-spur ratio of 23 dB (neglecting the laser). Referring back to Fig. 3 in the full-bias mode path, two cascaded athermal FBG filters, FB1 and FB2, are used to provide greater than 50 dB rejection of the laser reference. The pass band loss of each FBG filter is approximately 1 dB. Each FBG has a reflection bandwidth of 0.5-nm centered about 1556.21-nm and prevents operation in full-bias mode for output frequencies less than 63 GHz. Isolators IS4 and IS2 are necessary to absorb the reflected laser power.

Simulated data evaluated in Matlab using (2) to describe the MZM optical output power is provided in Fig 5. The ratio of the carrier frequency, $f_0$, to the modulating frequency, $f_m$, is approximately the same as the empirical data shown in Fig 4. The narrow line widths of this simulation more closely represent the true line widths because it is not restricted to the minimum resolution bandwidth imposed by the spectrum analyzer. Fig. 5(a) shows the simulated output in null-bias mode with $\theta_B = \pi$ radians and results in idealized suppression of the 0th and 2nd harmonic, with the two desired 1st harmonic tones 31 dB above the undesired 3rd harmonic tones. Fig. 5(b) shows the output in full-bias mode with $\theta_B = 0$ radians and results in idealized suppression of the 1st and 3rd harmonics. The two desired 2nd harmonic tones are 20 dB below the undesired carrier. The phase modulation depth, $\delta$, was set to 0.8 for both simulations.

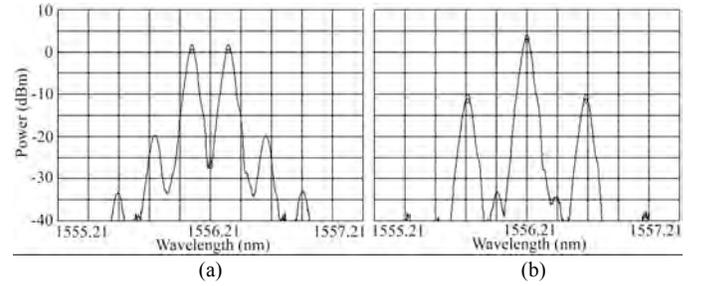

(a)                    (b)

Fig. 4. MZ1 optical output spectra. (a) Null-bias mode with 15.0 GHz reference input, the two desired 1st harmonic tones are +0.8 dBm and separated by 0.2423-nm (30.0 GHz). Note the suppression of the 1556.21-nm laser. (b) Full-bias mode with 24.0 GHz reference input, the two desired 2nd harmonic tones are -11.0 dBm and are separated by 0.7755-nm (96.0 GHz). Data was captured using an Agilent 86146B optical spectrum analyzer over a span of 2-nm (247.6 GHz) with a resolution bandwidth of 0.06-nm (7.4 GHz). The broad line widths are due to the large resolution bandwidth (minimum setting) of the spectrum analyzer.

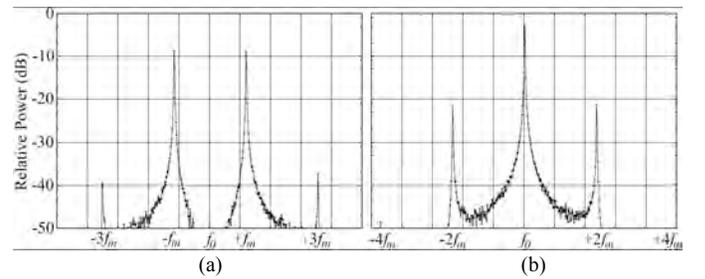

(a)                    (b)

Fig. 5. Matlab simulation of optical output power. (a) Null-bias mode with primary optical tones separated by $2f_m$, note the >40 dB suppression of the $f_0$ optical carrier relative to the desired sidebands. (b) Full-bias mode with primary optical tones separated by $4f_m$, note the desired sidebands are ≈ 19 dB below the undesired $f_0$ optical carrier.

### B. Optical Output Circuitry

An EDFA with a variable gain range of approximately 15-30 dB is used to compensate for component losses and to produce optical levels of approximately +4 dBm per tone at output connectors J2 and J3 (refer to Fig 3). In addition to providing gain, the EDFA is used to compensate for MZ1's reduced modulation efficiency at higher drive frequencies as indicated by the dotted lower trace of Fig. 6. The higher MZ1 output level for null-bias mode is partially compensated by the CP3 10 dB coupling loss. The EDFA gain is controlled to produce an optical output power variation of less than 0.5 dB peak-to-peak over the entire tuning range as seen in the solid upper trace in the figure. A gain versus frequency table is empirically derived in 1 GHz increments similar to that of the DC bias table for MZ1 port C. Interpolation is used to set the EDFA gain for any frequency within the valid tuning range.

The effect of the EDFA's amplified spontaneous emission (ASE) can be seen as a noise pedestal in Fig 7(a). This particular EDFA has a gain dependent noise figure ranging from approximately 1.5 to 4 dB at 1556.21-nm with the shape of the noise pedestal approximating the gain shape of the EDFA. Though the EDFA appears to add a significant amount of noise, nearly all of it is removed with the use of a 4-nm wide optical bandpass filter, FL3, shown in Fig 7(b).

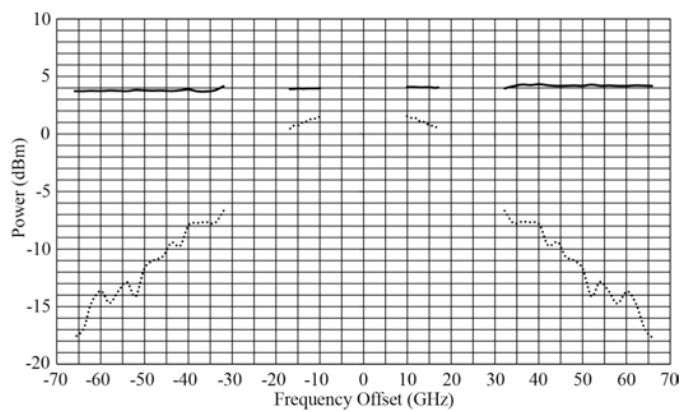

Fig. 6. Optical output signal leveling using EDFA gain control. Horizontal axis represents optical tone offset from laser center frequency of 192.642702 THz. The MZ1 output and the post leveled J2 connector output are represented by the dotted and solid traces, respectively.

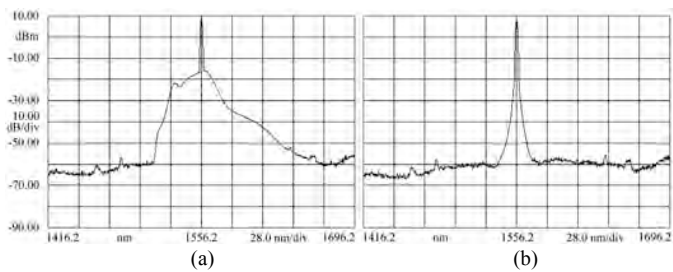

Fig. 7. Removal of EDFA ASE noise pedestal. (a) ASE noise pedestal generated by the EDFA, and (b) after the addition of the 4-nm wide BPF FL3. Both plots are with the laser synthesizer setup for 27 GHz output, the two optical tones appear as one because of the limited resolution bandwidth of the spectrum analyzer at a span of 28-nm/div.

## C. Mechanical Packaging

The MZM laser synthesizer is housed within a rack mountable electromagnetic interference (EMI) sealed chassis with dimensions of 19.0 x 8.75 x 22.0 inches, excluding handles and connectors. Total weight with covers is 42 pounds and AC power consumption is 58 W. The bottom and top views are shown in Fig. 8 and Fig. 9. Two 48 cubic-feet/minute cooling fans mounted on the lower portion of the rear panel draw air into the unit and provide cooling along the lower portion of the fixed aluminum deck. The airflow wraps around the gap near the front panel and flows across the upper deck in the reverse direction. Two EMI screened exhaust vents are provided on the upper portion of the rear panel. Thermal sensors are used to monitor the following temperatures: intake air, exhaust air, AR1, AR2, PS1 (+12 VDC supply), +9 V regulator, and the center of the fixed aluminum deck.

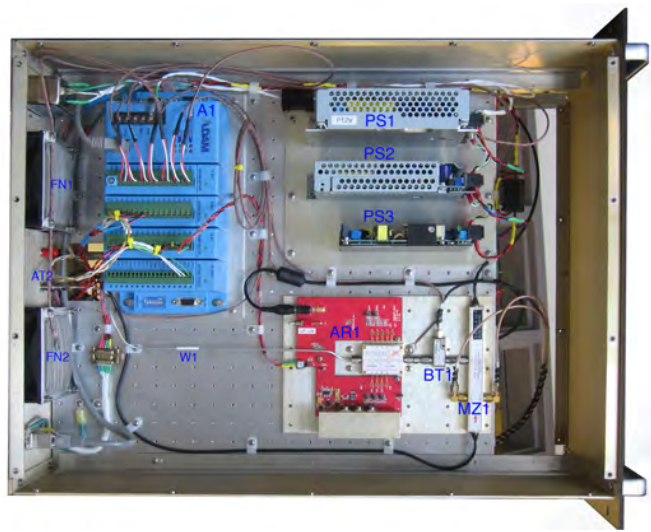

Fig. 8. Bottom view of MZM laser synthesizer chassis. PS1, PS2, PS3 are DC power supplies, A1 is the Ethernet controller. AR1, BT1, and MZ1 are the RF power amplifier, bias-T, and Mach-Zehnder modulator. Fans FN1 and FN2 draw air in from the rear panel.

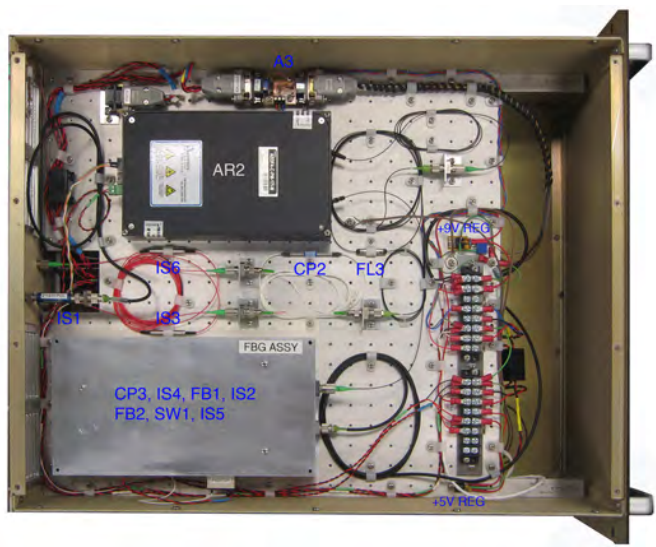

Fig. 9. Top view of MZM laser synthesizer chassis. The FBG assembly houses optical components CP3 through IS5. Output components AR2 through IS3/IS6 are mounted on the aluminum deck. Note the input isolator IS1 is mounted directly onto the rear panel.

## IV. Performance Characterization

### A. Optical Performance

The optical output performance is shown in Fig. 10 for the low and high operating frequency limits. It is apparent from the figure that the optical noise floor has increased for the 124 GHz case due to the input signal power reduction which requires an increase in the EDFA gain to maintain an output of +4 dBm per tone. The measured optical signal-to-noise ratios (OSNR) for the two plots in Fig. 10 are 33 and 18 dB for 26 and 124 GHz, respectively, and easily meets the ALMA requirement of 10 dB. The 0.5 dB tone imbalance seen in the right plot is a result of a small gain slope in the EDFA response (refer to Fig. 7(a)) and is well within the 3 dB requirement. At frequencies beyond 136 GHz, the EDFA reaches its maximum gain and the unit can no longer maintain the required optical output power of +3 dBm per tone.

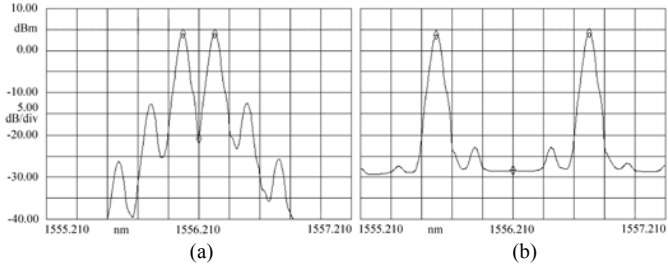

Fig. 10. Optical output spectra. (a) output tones for 26 GHz optical spacing, LSB/USB = 4.0/4.0 dBm. (b) output tones for 124 GHz optical spacing, LSB/USB = 3.7/4.2 dBm. Note the higher noise floor for the 124 GHz setting resulting from lower input signal power and higher required EDFA gain.

Fig. 11 represents the optical output power measured at J2 using an Oz Optics POM-300-IR power meter over a 12-hour duration for two different LO frequencies. The minimum specified output level is +3 dBm per tone with a stability requirement of ≤ 0.4 dB RMS.

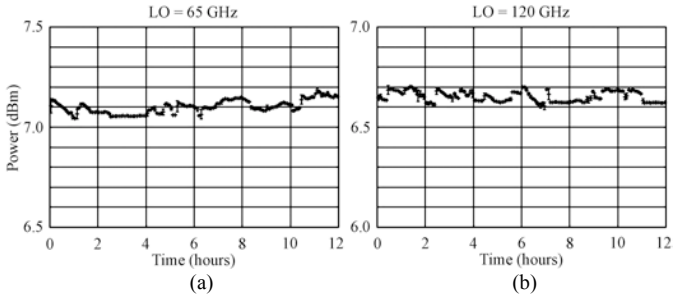

Fig. 11. Optical output power stability. Both (a) and (b) were measured with an optical power meter over a contiguous 12-hour duration. Peak-to-peak variation is 0.15 dB for 65 GHz, 0.12 dB for 120 GHz. The ALMA requirement is ≤ 0.4 dB RMS.

Fig. 12 represents the polarization extinction ratio (PER) measured at J2 using an Oz Optics ER-100-1290/1650-ER=40 PER meter over a 12-hour duration for two different frequencies. Note that the laser synthesizer unit has internal isolator/polarizer pigtail devices, IS3 and IS6, (refer to Fig. 3 and Fig. 9) installed just prior to output connectors J2 and J3 to ensure adequate PER performance. The apparently large PER variation exhibited in Fig. 12(a) may have been due to either or both the internal and external Panda fibers interfacing from IS3 to J2 to the PER meter. A PER of 30 dB translates to 1000:1 power ratio between the slow and fast axis and is extremely sensitive to microscopic movements and settling of the fibers. For PER measurements > 25 dB it is common to see an initial drop in PER values just after making the connection due to minute heating of the fiber from the optical signal itself.

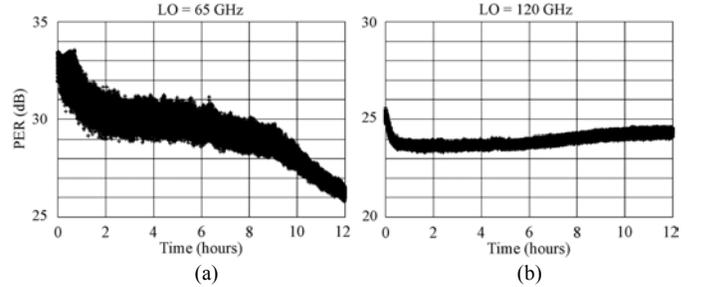

Fig. 12. Optical PER stability. Both (a) and (b) were measured with an optical PER meter over a contiguous 12-hour duration. Worst case PER is 25.8 dB for 65 GHz, 23.2 dB for 120 GHz. The ALMA requirement is ≥ 20 dB. The two measurements were taken on separate days.

### B. RF Performance

The RF test setup and description used to characterize the RF performance of the MZM laser synthesizer is shown in Fig. 13. The RF synthesizer is an Agilent E8257D with low phase noise option. The TeraXion laser reference provides a narrow 5 kHz wide (-3 dB bandwidth) polarized optical tone at a wavelength of 1556.21-nm. All optical interfaces are of FC/APC type and are keyed to the slow axis. Example optical spectra from the J2 unit output connector are shown in Fig. 10. The optical output is detected by a high performance W-band NTT IOD-PMW-09001-0 photomixer which accepts a nominal input wavelength of 1550-nm and provides output operation over 70-110 GHz via a WR-10 waveguide interface. This photomixer requires a low noise DC bias of -2.0 V and provides a responsivity, $r$, of 0.43 A/W. The RF power generated by the photomixer is expressed as

$$P_{RF} = 2(I_{PH})^2 R_L,$$ (6)

where $P_{RF}$ is the W-band RF output power, $I_{PH} = rP_{opt}$ is the average DC current generated by optical signal, and $R_L$ is the 50 Ohm load impedance [19]. The optical power, $P_{opt}$, into the photomixer is nominally 2.5 mW (+4 dBm) for each tone and corresponds to a photomixer current, $I_{PH}$, of 1.075 mA. Evaluating (6) yields a $P_{RF}$ value of 0.12 mW (-9.37 dBm) and is fairly close to the measured value of 0.10 mW (-10 dBm) shown in Fig. 14(b).

The Agilent 11970W harmonic mixer interfaces directly to the photomixer waveguide output and provides operation over 75-110 GHz. An Agilent 8563E RF spectrum analyzer outfitted with a phase noise measurement utility option operates in conjunction with the harmonic mixer, providing a 1st LO reference to the harmonic mixer for down conversion to an intermediate frequency (IF).

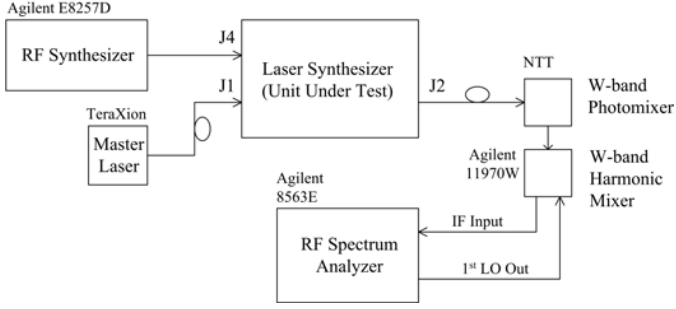

Fig. 13. Setup for RF Performance Tests. A low phase noise RF synthesizer and narrow line width laser are used to stimulate the unit under test. A W-band NTT photomixer is used to detect the difference frequency between the two light wave tones and provides the signal to the W-band harmonic mixer and RF spectrum analyzer.

A comparison of the spectral outputs for the 25.0 GHz input signal and 100 GHz photomixer output are shown in Fig. 14(a) and (b), respectively. In this case, the MZM laser synthesizer is operated in full-bias mode where the output frequency is four times the input reference frequency. Note the approximately 10 dB increase in noise floor for the 100 GHz signal. Power spectral density plots for the two signals are provided in Fig. 15. Since these plots only show the power spectrum from 3 kHz to 3 MHz, they exclude lower frequency contributions from flicker frequency noise, $1/f^{1}$, and random walk frequency noise, $1/f^{2}$. The most significant contribution seen in the plots is white frequency or random walk of phase noise, $1/f^{0}$ [20]. The total phase noise of the photomixer output shown in Fig. 15(b) consists of a number of individual contributors that are evaluated below.

The dark current and Shot noise [19] can be expressed as

$$P_{dark+Shot} = 3e(I_D + I_{PH})BR_L,$$   (7)

where, $I_D$ is the dark current ($\approx$ 1 nA, negligable), $e$ is the electron charge ($1.60\times10^{-19}$ C), and B is the bandwidth (35 GHz, limited by harmonic mixer).

The zero point noise [21] becomes significant at optical frequencies and can be expressed as

$$P_{ZP} = 2hfrI_{PH}BR_L,$$   (8)

where $h$ is Planck's constant ($6.63\times10^{-34}$ Js) and $f$ is the laser frequency (192.6 THz).

The Johnson noise [19] contribution of the photomixer is expressed as

$$P_{thermal} = 4kTB,$$   (9)

where $k$ is Boltzmann's constant ($1.38\times10^{-23}$ J/K), and $T$ is the ambient temperature ($\approx$ 300 K).

Finally, the contribution of the laser's relative intensity noise (RIN) [19] on the photomixer can be expressed as

$$P_{RIN} = (I_{PH})^2(RIN)BR_L,$$   (10)

where the value of RIN is estimated to be -155 dBc/Hz ($3.16\times10^{-16}$ Hz$^{-1}$).

Evaluating the quantities for (7)-(10) yields $P_{dark+Shot} = 9.03\times10^{-7}$ mW, $P_{ZP} = 2.07\times10^{-7}$ mW, $P_{Johnson} = 5.80\times10^{-7}$ mW, and $P_{RIN} = 6.39\times10^{-7}$ mW. The noise sources are incoherent and summing produces a $P_N$ value of $2.33\times10^{-6}$ mW. In comparison, the measured signal power, $P_{RF}$, exiting the photomixer is 0.1 mW. These values of signal and noise are converted to voltage using

$$V_{RMS} = \sqrt{P/50},$$   (11)

where $V_{RMS}$ is the RMS voltage, $P$ is power provided in Watts, and 50 is the characteristic impedance in Ohms. Using (11), the calculated signal voltage is $1.41\times10^{-3}$ V$_{RMS}$ and the noise is $6.83\times10^{-6}$ V$_{RMS}$. The phase noise contributed by $P_N$ can be calculated by

$$\phi_N = \arctan\left(\frac{V_N}{V_{RF}}\right),$$   (12)

and results in 0.2775 degrees RMS. The total phase noise measured by the RF spectrum analyzer in Fig. 13 is represented by

$$\phi_{total} = \sqrt{\phi_{LS}^2 + (4\phi_{ref})^2 + \phi_N^2},$$   (13)

and

$$\phi_{LS} = \sqrt{\phi_{total}^2 (4\phi_{ref})^2 - \phi_N^2},$$   (14)

where $\phi_{LS}$ and $\phi_{ref}$ are the phase noise of the MZM laser synthesizer and reference (RF synthesizer), respectively. From Fig. 15, the measured values for $\phi_{total}$ and $\phi_{ref}$ are 1.1297 and 0.2656 degrees, respectively. Using (14) to solve for $\phi_{LS}$ yields a value of 0.2656 degrees RMS. This value represents the residual phase noise of the MZM laser synthesizer at 100 GHz over an integration bandwidth of 3 kHz to 1 MHz² and is approximately one fourth of the ALMA requirement of 0.97 degrees (1 kHz – 1 MHz integration bandwidth). The 1 kHz and 1 MHz integration bandwidth limits are derived from the largest expected fiber length disparity (> 20 km) between antennas, and the YIG PLL bandwidth (see Fig. 16), respectively.

---

² Desired integration bandwidth is 1 kHz – 1 MHz, however, the measurement was limited to 3 kHz to 3 MHz by the test instrument.

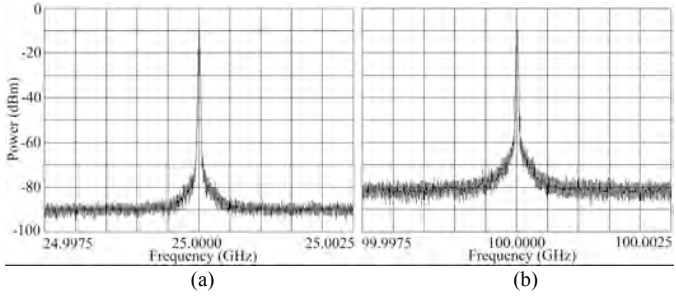

Fig. 14. Comparison of RF input and output spectra. (a) 25 GHz RF synthesizer, (b) 100 GHz photomixer output. Span is set to 5 MHz, resolution bandwidth 10 kHz, video bandwidth 1 kHz, for both plots.

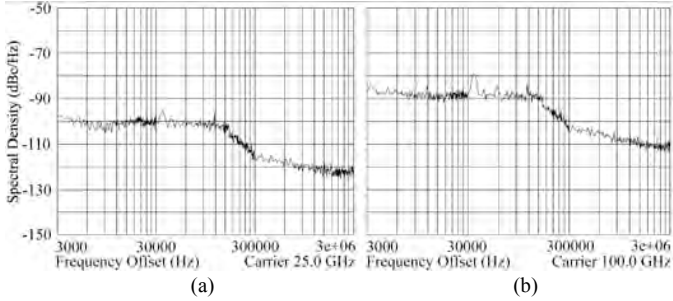

Fig. 15. Comparison of RF input and output spectral density. (a) 25 GHz RF synthesizer, (b) 100 GHz photomixer output. Integrated phase noise for the two plots are $\phi_{RF}$ = 0.2656 and $\phi_{total}$ 1.1297 degree RMS (3-3000 kHz), respectively.

## V. APPLICATION TO THE ALMA PROJECT

The MZM laser synthesizer was designed and constructed as an alternate to the baseline. As described in the previous section, the MZM laser synthesizer performance meets and exceeds the optical output stability, PER stability, and most significantly, the RF phase noise performance requirements. Fig. 16 describes the function of the laser synthesizer in relation to the overall central LO system [22].

The phase drift accumulated by the optical reference LO over the single-mode optical fiber is actively corrected by a real-time round trip phase correction system. The two tones from the laser synthesizer are transmitted through the line length corrector (LLC) to the antenna where they are reflected back on the orthogonal polarization of the fiber via a Faraday mirror. Each of the two optical tones are frequency shifted by the fiber frequency shifter (FFS) by 25 MHz in each direction, 50 MHz total. The photonic sub-array switch (PSAS) combines the transmitted and reflected wavelength domain LSB optical tones that are detected by a photomixer to produce a 50 MHz beat-note. This beat-note is phase compared to a 50 MHz system reference and is used to close the LLC servo control loop. The LLC utilizes a piezoelectric line stretcher that has the capability of tracking and compensating for fiber length changes of up to ±2-mm. This compensation range was determined by applying the thermal coefficient of expansion to the single-mode fiber located in buried conduit where thermal changes are expected to be minimal.

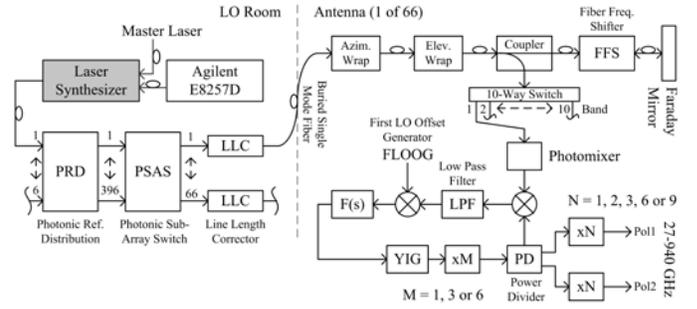

Fig. 16. Laser synthesizer in relation to central LO system. A pair of optical tones are generated by the laser synthesizer and transmitted to the antenna. The photomixer generates a RF difference frequency and is mixed with $Mf_{YIG}$ where $M$ is the multiplier number and $f_{YIG}$ is the YIG oscillator output frequency. The difference frequency between the photomixer and xM output is phased locked to the FLOOG reference. The photomixer and YIG PLL circuitry is combined into a packaged referred to as warm cartridge assembly (WCA). The final LO frequency is achieved after multiplication by N located outside of the WCA.

### A. Fast Frequency Switching Mode

Calibration for the ALMA telescope is performed frequently to obtain good quality images and reliable intensities and coordinates of sources in the sky. Band 3 operates over 84-116 GHz and is often used for calibration even if the actual science observations are performed using other bands. For example band 9, 602-720 GHz, where it is difficult to find suitable calibration sources. As a result, one of the requirements for the central LO system is the ability to switch frequency bands every 10 seconds, with a maximum switching time of 0.5 seconds.

One implication for switching LO frequencies for the MZM laser synthesizer is the non-flat optical loss versus frequency characteristic of MZ1 described previously in the dotted trace of Fig. 6. The EDFA gain is actively controlled to compensate for the optical loss differences between any two frequency states. Fig. 17 shows the results of the optical output power as measured with an optical power meter during a fast frequency switching operation. For the 65/120 GHz reference LO case, the uncorrected power imbalance between the two frequency states is 7 dB and is subsequently reduced to 0.4 dB using the EDFA gain control. The output power settles to within 90% of the final value within 0.2 seconds.

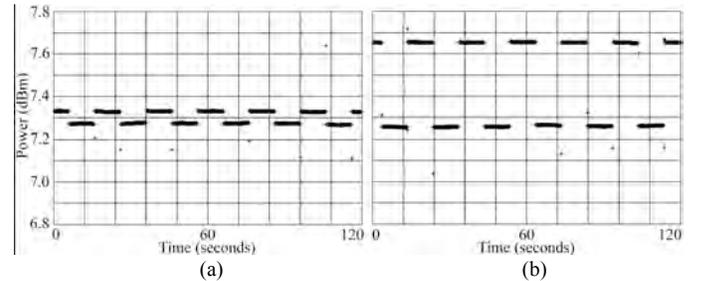

Fig. 17. Frequency switching mode optical output power versus time. (a) Frequency switching between 100 and 120 GHz in 10 second intervals, (b) switching between 65 and 120 GHz. Note the relatively small 0.05 and 0.40 dB power imbalance between the two frequency states for 100/120 GHz and 65/120 GHz, respectively.

### B. Line Length Corrector Compatibility

The ALMA central LO system employs an active phase correction system based on the round trip phase of the master laser tone of the baseline laser synthesizer. For the MZM laser synthesizer, however, the master laser is suppressed and the LLC must rely on the LSB tone for phase comparison. Under static frequency conditions the LLC locks reliably and tracks and corrects for the round trip phase of the LSB optical tone. During a switch in frequency, however, the LSB wavelength abruptly changes resulting in the LLC to lose lock. This loss of LLC lock during frequency switching events prevents calibration at different frequencies and is unacceptable for the ALMA project.

One solution to this issue is to offset the MZM laser synthesizer input reference laser wavelength by 5-nm and to reintroduce the master laser in addition to the two optical tones. Preliminary test results have shown this solution to be viable for the LLC, though it has yet to be determined what overall effect the third optical tone will have on the warm cartridge assembly (WCA) performance.

### C. Applied DC Bias Drift for MZM Device

The MZM laser synthesizer is operated in one of two operational modes, null-bias for bands 1 and 5 which require a photonic reference LO range of 27-34 GHz, and full-bias for the remaining bands which require a reference LO range of 65-122 GHz. Ports A and B of MZ1 are nominally set to 0 V and fine trimmed for intensity balance of the upper and lower optical paths in Fig. 1(a). The bias voltage presented to port C determines the mode, nominally +5 V ($V_\pi$) for null-bias mode and 0 V for full-bias mode. Long term operation with 5 V presented to port C causes a space charge accumulation near the electrodes. This in turn creates an electric shielding effect that requires the applied DC bias to be increased slowly with time in order to maintain suppression of the undesired optical tones. The phenomenon of DC bias drift is described in detail in [23]. If not compensated for, this phenomenon results in a runaway increase in the DC bias of port C and renders the MZM device unusable after a few weeks of operation.

One solution as implemented by [5] would be to replace the MZM with an optical phase modulator that does not require a DC bias. This approach, however, is not a feasible solution for our case since we need to operate in both null and full-bias modes for the wide frequency coverage, and in addition, requires the application of fine DC bias compensation.

Another proposed solution [12] is the use of a feedback control loop. This is accomplished by periodically measuring the power of the beat signal of the desired tone versus the undesired tone for both the null and full-bias modes. The DC bias value is dithered to determine the new DC bias value which results in a minimum beat-note power. A DC offset correction obtained in this fashion at a mid-band representative frequency in each mode can be applied to the entire bias table for all the modulating frequencies. This solution, however, will not be viable over long term continuous operations because of the runaway DC bias voltage effect described previously.

A simple remedy for the DC bias drift issue that we are presently utilizing is to reverse bias MZ1 port C with -10 V as part of a calibration routine. This calibration routine, however, requires a substantial amount of time, approximately 1 hour of calibration per 2 hours of operation in null-bias mode. Fortunately, null-bias mode is only utilized for bands 1 and 5. A more elegant solution is to alternate the port C bias voltage with $V_\pi$ and $V_{-\pi}$ during alternate operations in null-bias mode. This approach would minimize the calibration down time of the MZM laser synthesizer.

## VI. CONCLUSION

An engineering model of the MZM laser synthesizer was successfully developed and was shown to meet and exceed the ALMA performance specification requirements. The MZM laser synthesizer has the advantage of fast and reliable tuning along with extreme frequency stability [12] quantified by the Allan standard deviation, which translates to exceptional phase noise performance. Though the phase noise performance of the MZM laser synthesizer is approximately one quarter of the specification limit, other contributing factors cause the overall final LO improvement in phase coherence to be less substantial. Reference [24] has estimated that the potential improvement in the astronomical signal coherence is negligible for most of the lower bands and a modest $\approx$ 3% improvement in coherence for bands 9 and 10. There is a significant $\approx$ 13% improvement for the proposed band-11 (> 1 THz) science for ALMA, which is currently in the planning stages.

### ACKNOWLEDGMENT

The authors would like to thank William Shillue, Yoshihiro Masui, Jason Castro, and Skip Thacker from the National Radio Astronomy Observatory, Charlottesville, Virginia for their technical support and the use of their facilities. R. Srinivasan thanks Dr. Elsa Garmire from Dartmouth Collage, Hanover, New Hampshire for her invaluable advice and suggestions on various aspects of the Mach-Zehnder modulator. And finally the authors thank the IEEE reviews for providing informative feedback for this paper.

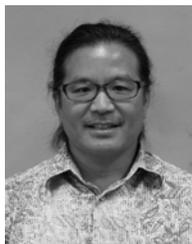

**Derek Y. Kubo** (M'10) received the B.S. degree in electrical engineering from the University of Washington, Seattle, WA, in 1984.

He was employed with Boeing Aerospace, Kent, WA, from 1984 to 1988, providing RF circuit board designs for JTIDS military radio testbed hardware. He was with NEC, Hawthorne, CA, as a Cell Site Engineer from 1988 to 1999. From 1989 to 1998 he was employed by Northrop Grumman Corporation, Redondo Beach, CA, as Department Staff Engineer for the high-speed modem section where his primary role was the development of a PSK demodulator product line for wideband satellite data links. His work also involved the development of associated equipment including modulators, adaptive equalizers, and cross polarization cancellers. He continued work with Northrop Grumman as a consultant from 1998 to 2002. In 2002, he joined Academia Sinica, Institute of Astronomy and Astrophysics, Hilo, HI, as Senior Microwave Engineer, where he is involved with the development of instrumentation for radio astronomy.

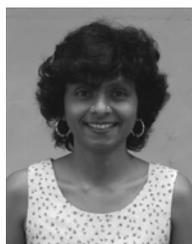

**Ranjani Srinivasan** received the B.Sc. degree in physics from Women's Christian College, Chennai, India in 1989, the M.Sc. degree in physics from the Indian Institute of Technology, Chennai, India in 1991, and the M.S. degree in astronomy from the University of Illinois at Urbana-Champaign, IL, in 2001.

She joined Academia Sinica, Institute of Astronomy and Astrophysics, Hilo, HI, in 2008 working for the Smithsonian Submillimeter Array interferometer telescope in software development as a Programmer. She was employed as a part time Science Instructor at the 'Imiloa Astronomy Center, Hilo, HI, from 2008 to 2010, and as a part time undergraduate Physics Instructor at the University of Hawaii, Hilo, HI, from 2009 to 2011. From 2009 to the present her interests have been focused on photonics based reference signals for the Atacama Large Millimeter Array interferometry telescope.

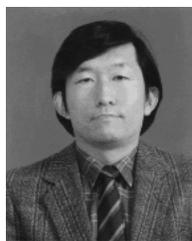

**Hitoshi Kiuchi** (M'09) received the B.E. and Ph.D. degrees in electronic engineering from university of electro-communications, Tokyo, Japan, in 1982 and 2001, respectively.

In 1982, he joined the Radio Research Laboratory, Ministry of Posts and Telecommunications (now the National Institute of Information and Communications Technology), Tokyo, Japan, where he was in charge of the VLBI (very long baseline interferometry) correlation processing system, data acquisition system, and the reference frequency system. In 2004, he joined the National Astronomical Observatory in Japan where his research activity is concerned with a photonic local oscillator system for the Atacama Large Millimeter Array interferometry telescope.

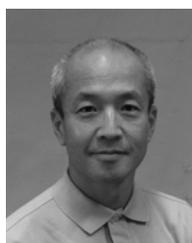

**Ming-Tang Chen** received his B.S. degree in physics from National Cheng Kung University, Tainan, Taiwan, in 1986, and M.S. and Ph.D. degrees in physics from University of Illinois at Urbana-Champaign, IL, in 1990 and 1993, respectively.

After his postdoctoral research at Case Western Reserve University, he joined Academia Sinica, Institute of Astronomy and Astrophysics, Taipei, Taiwan, in 1995 to work on instrumentation for the Smithsonian Submillimeter Array interferometry telescope and was dedicated for operation in 2003. His second instrumentation project in astronomy was the development and construction of the Array for Microwave Background

Anisotropy telescope, dedicated in 2006. He is currently serving as the project engineer representing Taiwan for the Atacama Large Millimeter Array telescope and is also involved in the development and deployment of a submillimeter telescope in Greenland. In addition to system integration, his expertise is with cryogenic techniques, quasi-optics, and millimeter wave instrumentation.